PAPER • OPEN ACCESS

# Parametric Investigation on Different Bone Densities to avoid Thermal Necrosis during Bone Drilling Process

To cite this article: Md Ashequl Islam *et al* 2021 *J. Phys.: Conf. Ser.* **2051** 012033

View the article online for updates and enhancements.

## You may also like

- Drilling of bone: thermal osteonecrosis regions induced by drilling parameters
  Mohd Faizal Ali Akhbar and Ahmad Razlan Yusoff

- Design Improvements in Conventional Drilling Machine to Control Thermal Necrosis during Orthopaedic Surgeries
  R V Dahibhate, S B Jaju and R I Sarode

- Performance of micro-drilling of hard Ni alloys using coated and uncoated WC/Co bits
  Philip A Primeaux, Bin Zhang and W J Meng

# Parametric Investigation on Different Bone Densities to avoid Thermal Necrosis during Bone Drilling Process

**Md Ashequl Islam**[1,1], **Nur Saifullah Kamarrudin**[1,2], **M.F.F. Suhaimi**[1,2], **Ruslizam Daud**[1,2], **Ishak Ibrahim**[1,2], **Fauziah Mat**[1,2]

[1]Faculty of Mechanical Engineering Technology, Universiti Malaysia Perlis, 02600 Arau, Perlis, Malaysia
[2]Centre of Excellence Automotive & Motorsports (MoTECH), Pauh Putra Campus, Universiti Malaysia Perlis (UniMAP), 02600 Arau, Perlis, Malaysia

saifullah@unimap.edu.my

**Abstract**. Bone drilling is a universal surgical procedure commonly used for internal fracture fixation, implant placement, or reconstructive surgery in orthopedics and dentistry. The increased temperature during such treatment increases the risk of thermal penetration of the bone, which may delay healing or compromise the fixation's integrity. Thus, avoiding penetration during bone drilling is critical to ensuring the implant's stability, which needs surgical drills with an optimized design. Bovine femur and mandible bones are chosen as the work material since human bones are not available, and they are the closest animal bone to human bone in terms of properties. In the present study, the Taguchi fractional factorial approach was used to determine the best design of surgical drills by comparing the drilling properties (i.e., signal-to-noise ratio and temperature rise). The control factors (spindle speed, drill bit diameter, drill site depth, and their levels) were arranged in an L9 orthogonal array. Drilling experiments were done using nine experimental drills with three repetitions. The findings of this study indicate that the ideal values of the surgical drill's three parameters combination (S1D1Di2) and their percentage contribution are dependent on the drilling levels of the parameters. However, the result shows that the spindle speed has the highest temperature effect among other parameters in both (femur and mandible) bones.

## 1. Introduction

Heat generation during bone drilling operations is a severe challenge in orthopedic and dental surgery. Although the bone is a self-repairing structural material, bone loss or bone fracture is one of the most common public health concerns due to automobile and motorcycle accidents, sports injuries, falls from a height, aging population, and other diseases [1]. Penetration of surgical tools may induce mechanical processes (sawing, plane cutting, drilling, grinding, etc.) and thermal damage in a bone [2]. The drilling parameters, mainly drill speed, feed rate, applied force, drilling depth, drill bit geometry, and bone damage, are directly related [3]. However, high temperature cutting in bone drilling can induce hyperthermia and even carbonization, leading to bone cell death and alteration in the bone property. In the conventional way of bone, drilling is still primarily conducted using hand drills, which means a blind operation with unknown hole depth and a feed rate manually controlled by the surgeons [4].

[1]

A mathematical model was used and compared the outcomes with experimentally measured temperature in cortical bone [5]. The computational result agreed with the experimental data. Jun Lee proposed a thermal model to investigate optimal drill bit geometries and machine conditions, which reduce thermomechanical damage during drilling [6]. Additionally, an experiment with a diamond-coated hollow bone drilling tool managed to minimize temperature rise during bone drilling successfully [5]. Optimization of drilling parameters has been conducted using Taguchi methodology by Pandey and others [7]. In the present study, the Taguchi fractional factorial approach was used on the femur and mandibular bones to determine the best combination of surgical drilling parameters by comparing the drilling outcomes (i.e., signal-to-noise ratio and temperature rise). The control factors (spindle speed, drill bit diameter, drill site depth, and their levels) were arranged in an L9 orthogonal array.

## 2. Selection of parameters

The drilling specifications were chosen based on prior research and the recommendations of experienced orthopedic surgeons. Parameters and levels are fixed for both femur diaphysis and mandibular bone drilling experiment. Table 1 contains a list of the factors investigated, along with their associated levels.

Table 1 Experimental Parameters with Levels

| No. | Factor | Units | Level 1 | Level 2 | Level 3 |
|---|---|---|---|---|---|
| 01 | Speed (S) | rpm | 10000 | 15000 | 20000 |
| 02 | Depth (D) | mm | 4 | 6 | 8 |
| 03 | Diameter (Di) | mm | 1 | 2 | 3 |
| Fixed Parameters | | | | | |
| 01 | Force (N) | 10.3 | | | |
| 02 | Feed (mm/min) | 20 | | | |

## 3. Methodology of Taguchi

Taguchi approach is a strong optimization strategy that increases process performance with the least number of tests. The Taguchi approach considers both controllable (control factors) and uncontrollable (noise factor) variables when determining the optimal combination of design parameter values to ensure that the actual result or operation is insensitive to noise factors. In this study, speed, depth, and diameter with the levels (1,2 and 3) are to be tested further experimentally. As given inTable 2, the L9 orthogonal array was chosen for the experimental purpose because it has eight degrees of freedom and can handle three tiers of four parameters. There are two ways of determining the S/N ratio. If the target function is to obtain the smallest possible value for the experimental findings, the smaller-the-better (Eq.1) principle is applied. Furthermore, the aim function is to maximize the value of experimental results; the greater-the-better (Eq.2) principle is applied.

$$\frac{S}{N} = -10 \; log\left(\frac{\sum Y^2}{n}\right) \qquad (1)$$

$$\frac{S}{N} = -10 \; log\left(\frac{\sum(\frac{1}{Y^2})}{n}\right) \qquad (2)$$

## 4. Experimental design and setup

This experiment utilized the drill bits with diameters of 1mm, 2mm, and 3mm manufactured of HSS-6542 135° with a typical cutting point of head geometry. The trials are conducted at 10000, 15000, and 20000 revolutions per minute with a drilling depth of 4, 6, and 8mm. Before commencing the drilling procedure, the bone was measured using the vernier caliper scale. A drill bit with a diameter of 1.2mm is used to pre-drill the bone to place thermocouples into the bone. As a result, the gap between the drilling holes midpoint and the thermocouple pre-drilled holes midpoint would be fixed at a 1mm distance. A T-type thermocouple was connected with a multi-channel temperature meter, and the reading data was stored in a pen drive attached with a meter USB port. Later, the data was analyzed by using Minitab software. Bovine femur and mandible bones are chosen as the work material since human bones are not available, and they are the closest animal bone to human bone in terms of properties. The bovine femoral and mandibular bones were received shortly after slaughter from a nearby slaughterhouse. The experiments were carried out within a few days to preserve the mechanical and thermo-physical characteristics of the bone. The bone was preserved in the saline water under – 20ºC temperature. There were no animals killed explicitly for the sake of this study. Overall experimental design and setup are shown in Figure 1.

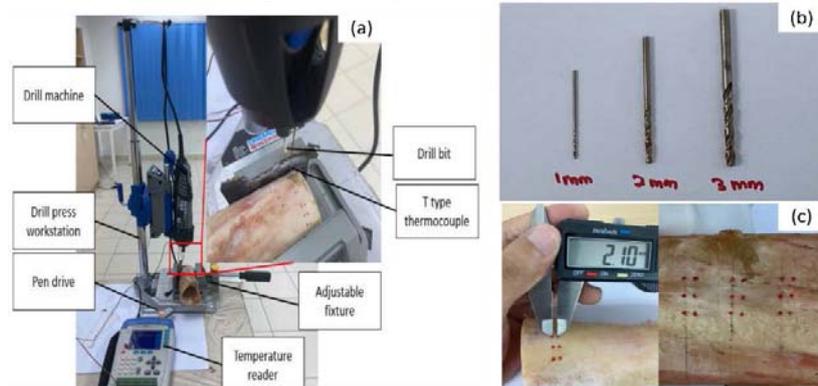

Figure 1 Overall Setup for Bone Drilling with the T-type Thermocoule inserted into the Bone, (b) High-Speed Steel Drill Bit, (c) Taking Measurement to Drill on the Bone with Precision

## 5. Results and analysis

The values obtained after every experiment for various output parameters have been recorded, as shown in Table 2. The control factors speed, depth, diameter, and the level of these three parameters were designed into nine experimental combinations. All nine combinations have experimented with 3 times repetition in order to reduce the experimental error, and the average values have been considered for analysis. To minimize bone necrosis, the heat generation should be as low as achievable so that the L9 experimental observations matrix is solved to optimize the parameters using Equation (1), smaller the better. The Minitab software analyzed collected data to get the optimized SN ratio (signal-to-noise) value for each experiment. The same process was repeated to generate the mandibular bone drilling experimental result.

Table 2 Femur and Mandible Bone Drilling Result from Taguchi L9 Experiments

| L9 Orthogonal array matrix | | | | Femur bone drilling | | Mandible bone drilling | |
| --- | --- | --- | --- | --- | --- | --- | --- |
| Exp. No. | Speed (rpm) | Depth (mm) | Diameter (mm) | Mean Temperature (ºC) | SNRA (dB) | Mean Temperature (ºC) | SNRA (dB) |
| 1 | 10000 | 4 | 1 | 35.6667 | -31.0487 | 31.7667 | -30.0409 |
| 2 | 10000 | 6 | 2 | 39.5000 | -31.9417 | 34.9667 | -30.8734 |
| 3 | 10000 | 8 | 3 | 39.7333 | -31.9873 | 35.9667 | -31.1215 |
| 4 | 15000 | 4 | 2 | 37.5667 | -31.5039 | 35.8333 | -31.0870 |
| 5 | 15000 | 6 | 3 | 43.1333 | -32.6974 | 39.8333 | -32.0151 |
| 6 | 15000 | 8 | 1 | 48.6000 | -33.7398 | 44.0333 | -32.9050 |
| 7 | 20000 | 4 | 3 | 53.6000 | -34.5839 | 42.0333 | -32.4856 |
| 8 | 20000 | 6 | 1 | 53.2000 | -34.5258 | 47.5667 | -33.5525 |
| 9 | 20000 | 8 | 2 | 51.0667 | -34.1721 | 46.6333 | -33.3816 |

The estimated S/N ratio and mean temperature for each parameter at different values are presented in Table 3 for femur bone and Table 6 for mandible bone. The optimal values from each parameters level are shown in bold numbers.

Table 3 Response Table for Heat Generation (Femur)

| Parameters | Signal-to-ratio response values | | | Mean temperature response values | | |
|---|---|---|---|---|---|---|
| | Level 1 | Level 2 | Level 3 | Level 1 | Level2 | Level 3 |
| Speed (rpm) | **-31.66** | -32.65 | -34.43 | **38.30** | 43.10 | 52.62 |
| Depth (mm) | **-32.38** | -33.05 | -33.30 | **42.28** | 45.28 | 46.47 |
| Diameter (mm) | -33.10 | **-32.54** | -33.09 | 45.82 | **42.71** | 45.49 |

Table 4 Response Table for Heat Generation (Mandibular)

| Parameters | Signal-to-ratio response values | | | Mean temperature response values | | |
|---|---|---|---|---|---|---|
| | Level 1 | Level 2 | Level 3 | Level 1 | Level 2 | Level 3 |
| Speed (rpm) | **-30.68** | -32.00 | -33.14 | **34.23** | 39.90 | 45.41 |
| Depth (mm) | **-31.20** | -32.15 | -32.47 | **36.54** | 40.79 | 42.21 |
| Diameter (mm) | -32.17 | **-31.78** | -31.87 | 41.12 | **39.14** | 39.28 |

The values from response Table 3 and Table 4 for heat generation are helpful to generate a response graph with respect to the level of each parameter. As shown in Figure 2 (a) the response temperature for femur bone drilling, (b) for mandible and (c) for comparison in one graph (femur & mandible), where the control parameters are spindle speed, drill depth and drill bit diameter with all three levels were plotted in the graph.

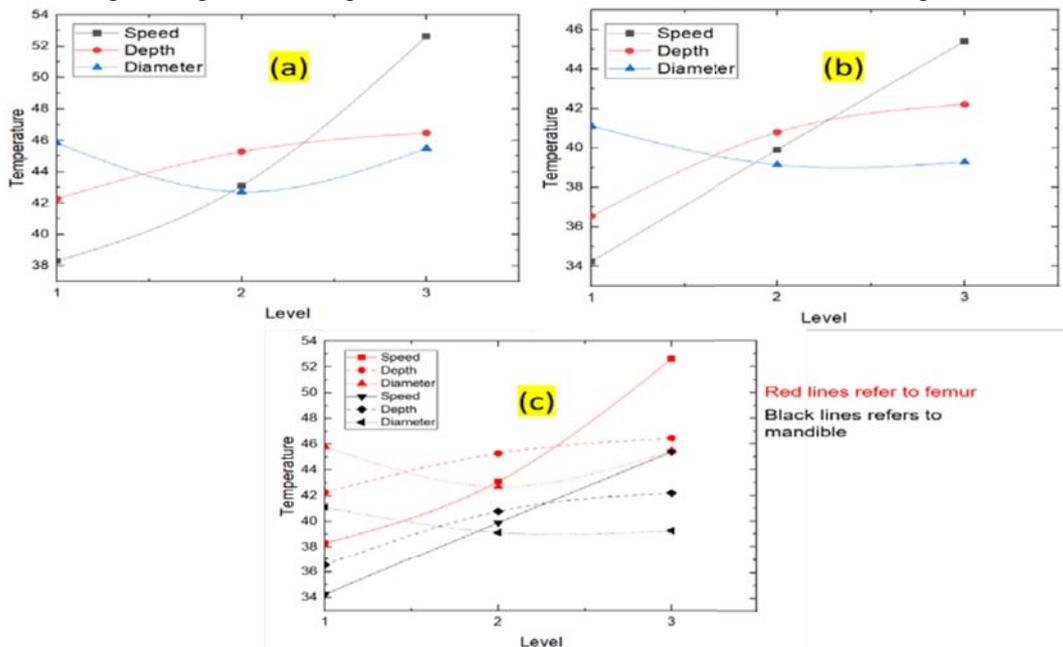

Figure 2  Mean Temperature Response of (a) Femur Bone, (b) Mandible Bone & (c) Comparison in one Graph

From graph (a), it is found that, during drilling of the femur bone, the parameter speed (38.30 ºC) in level 1, depth (42.28 ºC) in level 1, and diameter (42.71 ºC) in level 2 gave the optimized temperature. For graph (b) shows similar patterns of temperature rise, speed (34.23 º) in level 1, depth (36.54 ºC) in level 1, and diameter (39.14) in level 2 gave the lower heat elevations. Lastly, graph (c) shows the increment in the level of each parameter (expect drill diameter) experienced a heat elevation. A fast rotational speed transfers more energy from the spinning tool to the bone specimen, resulting in significant heat production at the drilling site. Hight of drill depth affects temperature too, because, with the increment of drill depth, drilling time and energy would increase too, leading to thermal necrosis.

### 5.1 Analysis of variance for femur

ANOVA statistic approach was used to analyze the data to determine the significant factors affecting the temperature rise. The temperature was calculated with a contribution of parameters at a 95% confidence interval. To highlight the impact of each parameter, the variation data for each component and their interaction were computed. The greater the contribution value, the higher the temperature rise for that parameter.

As shown in Table 5, the spindle speed (78.87%) has the greatest influence, followed by the depth of the drill site (9.12%) and the diameter of the drill bit (4.16%) had the less noise ratio output or generate less temperature. Here noise is our output parameter (temperature), based on given signals parameters (speed, depth, diameter) to the system.

Table 5 Analysis of Variance for SN Ratios (femur)

| Source | DF | Seq SS | Adj SS | Adj MS | F | P | Contribution |
|---|---|---|---|---|---|---|---|
| Speed (S) | 2 | 11.8067 | 11.8067 | 5.9034 | 10.05 | 0.091 | 78.87% |
| Depth (D) | 2 | 1.3652 | 1.3652 | 0.6826 | 1.16 | 0.463 | 9.12% |
| Diameter (Di) | 2 | 0.6229 | 0.6229 | 0.3114 | 0.53 | 0.654 | 4.16% |
| Residual Error | 2 | 1.1749 | 1.1749 | 0.5875 | | | 7.85% |
| Total | 8 | 14.9696 | | | | | 100% |

| | Predicted value | Experimental value | Error range |
|---|---|---|---|
| Optimal combination | S1D1Di2 | S1D1Di2 | ±0.4 |
| S/N ratio | -30.292 | -30.6551 | |

### 5.2 Analysis of variance for mandible

Table 6 has the effect on heat elevation during mandible bone drilling; spindle speed (75.52%) has the most significant influence, followed by the depth of the drill site (21.51%) and the diameter of the drill bit (2.10%).

Table 6 Analysis of Variance for SN Ratios (mandible)

| Source | DF | Seq SS | Adj SS | Adj MS | F | P | Contribution |
|---|---|---|---|---|---|---|---|
| Speed (S) | 2 | 9.1044 | 9.1044 | 4.55220 | 78.75 | 0.013 | 75.52% |
| Depth (D) | 2 | 2.5922 | 2.5922 | 1.29611 | 22.42 | 0.043 | 21.51% |
| Diameter (Di) | 2 | 0.2426 | 0.2426 | 0.12132 | 2.10 | 0.323 | 2.01% |
| Residual Error | 2 | 0.1156 | 0.1156 | 0.05780 | | | 0.96% |
| Total | 8 | 12.0549 | | | | | 100% |

| | Predicted value | Experimental value | Error range |
|---|---|---|---|
| Optimal combination | S1D1Di2 | S1D1Di2 | ±0.7 |
| S/N ratio | −29.305 | -30.0212 | |

To conclude, all factors substantially influence the amount of heat generated during bone drilling, but spindle speed has the greatest effect. Thus, it is advised that a lower spindle speed with a modest depth progression will be used to limit heat rise during orthopedic or dental surgery. It would generate less frictional energy over a shorter time period, preventing the drill bit's heat from spreading to nearby bone structures.

## 6. Conclusion

The purpose of this study was to examine the osteonecrosis of bone that occurs during orthopedic and dental bone drilling utilizing the CBD technique. A comparative experimental research in different bone densities was conducted to see which factor combinations can assist to minimize osteonecrosis during bone drilling using Taguchi method. The following findings have been drawn through experimental observation and comparative research.

- In the given parameter range, the rotational speed was the main factor responsible for heat generation. It contributes 78.87% to the femur bone drilling result (see Table 5) and 75.52% to the mandible bone drilling result (Table 6).
- The femur has a high bone density in contrast to the mandible. The temperature gets elevated when the bone has density; it requires more energy and time to drill.
- Speed and depth in the mandibular experiment significantly affect the outcome (heat generation) since each parameter has a P value less than 0.05 (Table 6).
- According to Taguchi and experimental results shown from our designed experiment, S1D1Di2 has the best combination for bone drilling in all types of densities. Moreover, the proposed combination generates the least amount of heat during femur and mandibular bone drilling.
- The confirmation test agreed with the experimental results within an error range of ±0.4 for the femur and ±0.7 for the mandible.

## 7. Acknowledgements

The author would like to acknowledge the support from the Fundamental Research Grant Scheme (FRGS) under a grant number of FRGS/1/2018/TK03/UNIMAP/02/8 from the Ministry of Higher Education Malaysia.